**Forecasting Political Stability in GCC Countries**

**Mahdi Goldani**

m.goldani@hsu.ac.ir

**Abstract:**

Political stability is crucial for the socioeconomic development of nations, particularly in geopolitically sensitive regions such as the Gulf Cooperation Council Countries: Saudi Arabia, UAE, Kuwait, Qatar, Oman, and Bahrain. This study focuses on predicting the political stability index for these six countries using machine learning techniques. The study uses data from the World Bank's comprehensive dataset, comprising 266 indicators covering economic, political, social, and environmental factors. Employing the Edit Distance on Real sequence (EDR) method for feature selection and XGBoost for model training, the study forecasts political stability trends for the next five years. The model achieves high accuracy, with mean absolute percentage error (MAPE) values under 10%, indicating reliable predictions. The forecasts suggest that Oman, the UAE, and Qatar will experience relatively stable political conditions, while Saudi Arabia and Bahrain may continue to face negative political stability indices. The findings underscore the significance of economic factors such as GDP and foreign investment, along with variables related to military expenditure and international tourism, as key predictors of political stability. These results provide valuable insights for policymakers, enabling proactive measures to enhance governance and mitigate potential risks.

## 1. introduction

political stability has various definitions in previous studies. political stability encompassing categories such as civil wars, democratic reversals, genocides and politicides, and state collapse which are comparatively rare but extremely essential [1], [2]. Political instability, in addition to the obvious huge costs for countries, including human, political, and social costs, often results in huge economic losses for unstable countries and their neighbors, which are rarely compensated by post-destabilization growth [3]. Most political instabilities require multilateral and sometimes intense international responses.

Therefore, there has been considerable and significant interest in predicting periods of instability in countries and among researchers [4]. Forecasting political stability can be done by using analytical tools, data models, and historical trend analysis in order to estimate the events that will happen in the future and impact the governance and stability of a nation or region. One can test hypotheses regarding impending challenges towards political stability by analyzing economic performances, social unrest, and political violence against governance quality and international relations. For example, demographic pressures of a growing youth population, combined with high levels of unemployment, may increase the risk of social unrest, particularly if the government is unable to address popular grievances. Similarly, frail institutional capacity, when mixed with political corruption, can often heighten tensions and lead to different forms of political instability. This would thus allow governments and policymakers to adopt precautionary measures-for instance, reforms, social programs, or diplomatic initiatives-to mitigate risks and ensure stability through efficacious forecasting. Especially in fragile states, this will be particularly critical, where external shocks-possibly economic crises or natural disasters-are likely to destabilize political order.

The GCC Countries and its coastal areas are the world's largest single source of petroleum [5]. The GCC countries produce 30% of the world's crude oil and play crucial role in global energy markets. The GCC member states, including Saudi Arabia, Kuwait, the United Arab Emirates, Qatar, Bahrain, and Oman, in addition to Iraq and Iran, are the mainstays in the production of the commodity within the region. The world energy statistics demonstrate in December of 2023 the GCC countries produce 31% of world oil (fig.1). Therefore, any changes in regional politics are monitored by the world. One of the most important disturbances in the region is the Gulf War in 1991, which cost a lot to the region and the beneficiary countries [6]. Considering the importance of political stability in the GCC region, this study aims to predict the political stability index published by the world Bank for the next five years in these countries.

fig1. Global Oil Production by Region

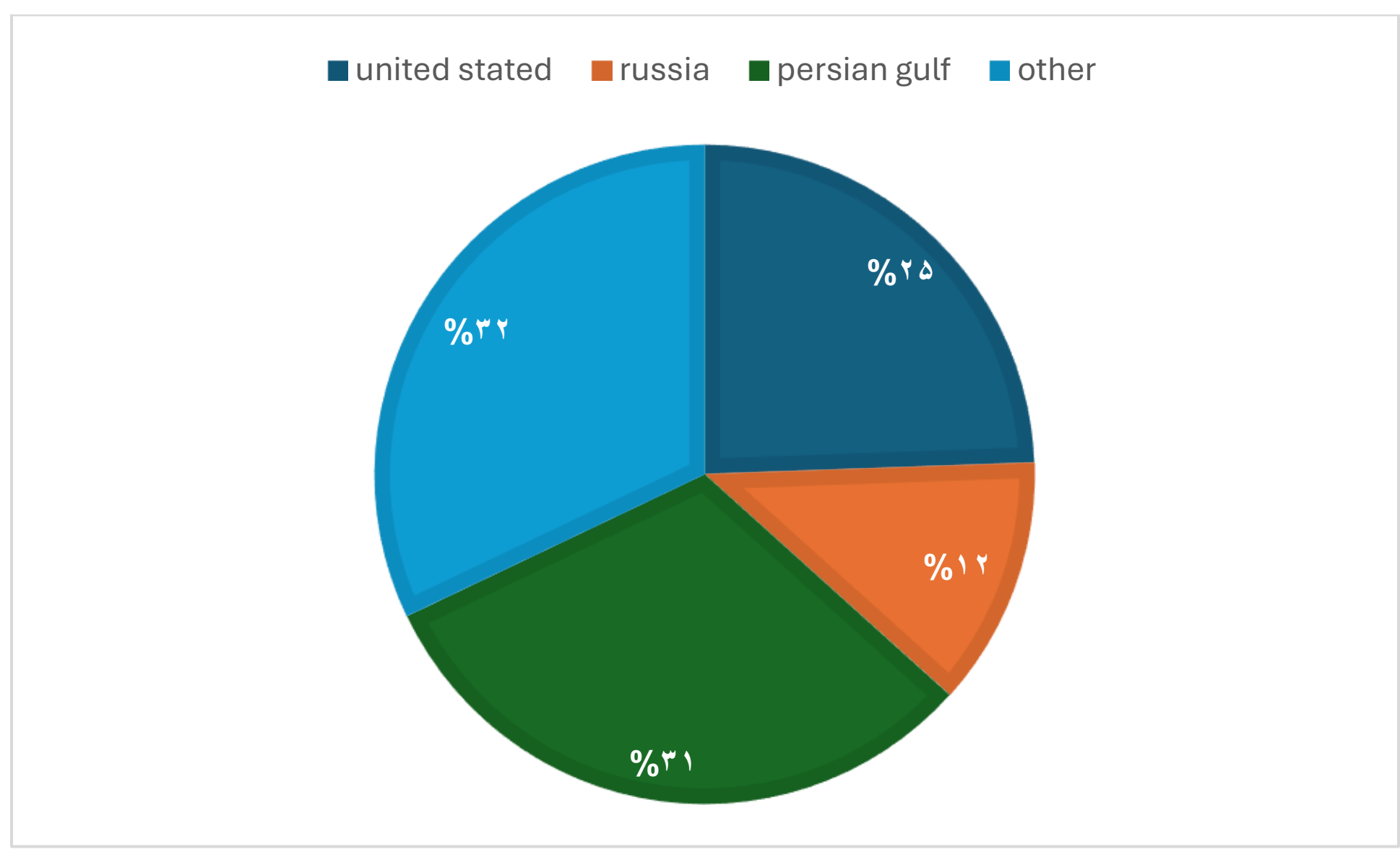

Source: world energy statistics 2023

After reviewing the studies published in the field, in order to obtain a comprehensive view of the factors affecting political stability and build a comprehensive model for better prediction, using a method of selecting the characteristics of 10 variables with the highest degree of relevance to political stability in the countries The subject of investigation is selected and then forecasting for the next 5 years is done using the XGBOOST method.

2. Literature review

The literature on political instability has focused on a range of predictive models using country-year data and macroeconomic factors. The diversity in predictors and methods highlights the ongoing debates within the field regarding the best indicators for forecasting instability. Goldani and Asadi [7] identified the similarities between the political stability index and the indicators presented in the World Bank for the two regions of the Caucasus countries and the OECD. the results show that economic and social factors are the most vital elements in political stability defining. Ulfelder [8] examined mass killings, using predictors like population size, GDP per capita, civil wars, and anocracy. His model, based on data from 1945 to 2011, was cross validated to ensure predictive accuracy. Another economic factor that is directly related to political stability is economic growth. Rashid and Rashid [9] showed that with a high propensity for government collapse, growth is significantly lower than otherwise. Bitar et al [10] investigated political stability as essential factors that determine foreign direct investment. Also, khan et al [11] showed that renewable energy development the stability of countries.

Besides the social and economic factors, environmental factors can be determinant. Kennedy's [12] work on political instability incorporated similar predictors to Goldstone, focusing on polity codes, infant mortality, and conflicts in neighboring states. His model covered data from 1955 to 1994 and extended predictions to 2012. Akhbarizadeh et al. [13]. The potential for bioaccumulation and biomagnification of pollutants could lead to public health crises, undermining trust in institutions tasked with protecting citizens. As environmental degradation intensifies, the potential for political instability rises, necessitating comprehensive forecasting models that consider ecological health as a determinant of political stability. Baillie et al [2] proposes two minimal logistic regression models for predicting political instability, utilizing polity code, infant mortality, and years of stability as predictors. The models prioritize interpretability and demonstrate competitive predictive power, validated through Monte Carlo simulations, making them more useful for policymakers due to their explainable outcomes. Gimpel et al. [14] provide insights into the urban-rural divide in political behavior, which may further exacerbate tensions as marginalized populations seek greater representation and resource allocation.

Based on previous studies, economic variables such as investment and economic growth, exports, social variables including population growth rate, infant mortality rate, pollution and political variables such as the type of regime are known as influencing variables on political stability. As an innovative move, this study uses the World Bank data set as a comprehensive data set in all economic, environmental, social and political fields, instead of being satisfied with the research literature, and thus the variables that determine political stability. from this dataset based on a feature selection method.

## 3. Dataset and methodology

### 3-1. dataset

The World bank reports 266 indicators for 217 countries and 49 regions in each year. Indicators illuminate the economic, social, political and environmental circumstance in each country and area therefore, it considers as a comprehensive dataset. In this study, all indicators of the World Bank are used to find out which indicators resemble political stability and are good explainers of target value.

Table1. features of datasets

| **COUNTRY** | | | | | | | | | | |
|---|---|---|---|---|---|---|---|---|---|---|
| **BAHRAIN** | International tourism , | Households and NPISHs Final | Service imports (BoP, | Net primary income (Net | Changes in inventories | Gross fixed capital formatio | Arms imports (SIPRI trend | Foreign direct investment, net | Stocks traded, total value | International tourism, receipts |

| | expenditures for travel items (current US$) | consumption expenditure, PPP (constant 2021 international $) | current US$) | income from abroad) (current US$) | (current US$) | n (current US$) | indicator values) | (BoP, current US$) | (current US$) | for travel items (current US$) |
|---|---|---|---|---|---|---|---|---|---|---|
| **UNITED ARAB EMIRATE** | Manufacturing, value added (constant LCU) | General government final consumption expenditure (current US$) | Agriculture, forestry, and fishing, value added (current US$) | International tourism, expenditures (current US$) | Services, value added (current US$) | Gross capital formation (current US$) | Arms imports (SIPRI trend indicator values) | Adjusted savings: consumption of fixed capital (current US$) | Foreign direct investment, net inflows (BoP, current US$) | Taxes on goods and services (current LCU) |
| **KUWAIT** | Households and NPISHs Final consumption expenditure (current US$) | Manufacturing, value added (current US$) | Households and NPISHs final consumption expenditure: linked series (current LCU) | International tourism, receipts for travel items (current US$) | Service exports (BoP, current US$) | Personal remittances, received (current US$) | Foreign direct investment, net inflows (BoP, current US$) | Military expenditure (current USD) | Portfolio equity, net inflows (BoP, current US$) | International tourism, expenditures for passenger transport items (current US$) |
| **QATAR** | GDP per person employed (constant 2021 PPP $) | Population, male | Urban population | Net migration | Fixed broadband subscriptions | Population in largest city | GDP per capita (constant LCU) | Air transport, registered carrier departures worldwide | Fixed telephone subscriptions | GNI per capita (current LCU) |
| **SAUDI ARABIA** | Exports of goods and services (current US$) | Merchandise imports (current US$) | Service imports (BoP, current US$) | Primary income receipts (BoP, current US$) | External balance on goods and services (current US$) | Changes in inventories (current US$) | Primary income payments (BoP, current US$) | International tourism, expenditures for travel items (current US$) | International tourism, receipts for travel items (current US$) | Foreign direct investment, net outflows (BoP, current US$) |
| **OMAN** | Taxes less subsidies on products (current US$) | Manufacturing, value added (constant LCU) | Final consumption expenditure (current US$) | Manufacturing, value added (current US$) | General government final consumption expenditure (current US$) | Net trade in goods and services (BoP, current US$) | Portfolio equity, net inflows (BoP, current US$) | Adjusted net savings, including particulate emission damage (current US$) | Changes in inventories (current US$) | Net official development assistance received (current US$) |

### 3-2. methodology

Machine learning (ML) is a powerful tool for making accurate and reliable predictions and, a sub-category of computational intelligence techniques mainly employed for deriving definitive information out of large sets of data for pattern recognition, classification, function approximation, and so forth. With the availability of vast datasets in the era of Big Data, producing reliable and robust forecasts is of great importance [15, 16, 17]. XGBoost is used in this study as an ensemble learning-based algorithm, where a set of base models are combined to create a model that obtains better performance than a single model [18]. It is considered a good method due to its combination of accuracy, efficiency, and flexibility. It surpasses in handling large datasets, managing missing data, preventing overfitting, and offering various tuning options. Its ability to process data quickly, handle imbalanced datasets, and provide interpretable models makes it a go-to choice for many machine learning practitioners. Whether for academic research or industrial applications, XGBoost remains a top choice for building robust and effective predictive models.

**3-2-1. Extreme gradient boosting (XGBoost)**

Chen and Guestrin [19] have created Extreme gradient boosting (XGBoost). XGBoost stands for "Extreme Gradient Boosting", where the term "Gradient Boosting" originates from the paper Greedy Function Approximation: A Gradient Boosting Machine, by Friedman [20]. XGBoost is based on the gradient boosting algorithm, a key method in ensemble learning. It combines weak classifiers to create a stronger model, enhancing efficiency and flexibility compared to a single model. By iteratively building decision trees, XGBoost improves classification performance.

A salient characteristic of objective functions is that they consist of two parts: training loss and regularization term Eq. (1)

$$obj^{(t)} = l(f_t) + \Omega(f_t) \quad (1)$$

In Eq. (1), $f_t$ representing the t-th tree model, $l(f_t)$ is the loss function in the risk prediction, $\Omega(f_t)$is the regular term used to reduce overfitting, which can be expressed as Eq.(2)

$$\Omega(f_t) = \gamma T + \frac{1}{2}\lambda\|\omega\|^2 \quad (2)$$

In Eq. (2), T represents the number of leaf nodes in the t-th decision tree, $\gamma$ and $\lambda$ can decide penalty strength together, $\omega$ representing the weight value on each leaf node. The training loss measures how predictive model is with respect to the training data. A common choice of $L$ is the mean squared error, which is given by Eq.(3)

$$l(f_t) = \sum_i (y_i^t - \hat{y}_i^t)$$
(3)

The prediction results of the model are the weighted sum of all the decision trees, when the t-th iteration is performed, the prediction result can be expressed by Eq(4).

$$\hat{y}_i^{(t)} = \sum_{k=1}^{k} f_{(x_i)=}\, \hat{y}_i^{(t-1)} + f_t(x_i),\ f_k \epsilon F \quad (4)$$

In Eq. (4), $f_t(x_i)$ represents the t-th tree model, F is the decision tree space, is also the set of all sample risk prediction decision trees. Here $\hat{y}_i^{(t)}$represents the prediction results of the sample I after the t-th iteration, and $\hat{y}_i^{(t-1)}$ represents the prediction results of the previous t-1 trees. Therefore, the objective function can be expressed as the formula for Eq. (5):

$$obj^{(t)} = \sum_{i=1}^{n} l\left(y_i, \hat{y}_i^{(t)}\right) + \Omega(f_t) + costant \quad (5)$$

Over fitting is common and undesirable machine learning methods behavior. Overfitting is the production of an analysis that corresponds too closely or exactly to a particular set of data and may therefore fail to fit additional data or predict future observations reliably. One of the solutions to avoid over fitting is tuning parameters to maximize model performance. There are a bunch of hyperparameter tuning methods. By considering the small dataset, Grid Search is a great option because it exhaustively explores all combinations of hyperparameters. The list of hyperparameters of XGboost is represented in table2.

Table2. hyperparameters of XGboost

| **HYPERPARAMETER** | **DESCRIPTION** | **TYPICAL VALUES** |
|---|---|---|
| **N_ESTIMATORS** | Number of trees (or boosting rounds). | 100, 500, 1000 |
| **LEARNING_RATE** | Step size shrinkage used to prevent overfitting. | 0.01, 0.1, 0.3 |
| **MAX_DEPTH** | Maximum depth of each decision tree. | 3, 5, 7, 10 |
| **MIN_CHILD_WEIGHT** | Minimum sum of instance weight (Hessian) needed in a child node. | 1, 3, 5 |
| **GAMMA** | Minimum loss reduction required to make a further partition on a leaf node. | 0, 0.1, 0.5, 1 |
| **SUBSAMPLE** | Fraction of samples used for building each tree. | 0.6, 0.8, 1 |
| **COLSAMPLE_BYTREE** | Fraction of features to be used for building each tree. | 0.6, 0.8, 1 |
| **COLSAMPLE_BYLEVEL** | Fraction of features used per tree at each level of boosting. | 0.6, 0.8, 1 |
| **LAMBDA (L2 REGULARIZATION)** | L2 regularization term to control model complexity and prevent overfitting. | 0, 1, 5, 10 |
| **ALPHA (L1 REGULARIZATION)** | L1 regularization term to control model complexity and sparsity. | 0, 1, 5, 10 |
| **SCALE_POS_WEIGHT** | Balancing of positive and negative weights for imbalanced classification. | 1, depending on class imbalance |

3-2-2. Model Evaluation

Due to the limited sample size, the cross-validation method is chosen to obtain reliable insights. Cross-validation is particularly useful for small datasets because it maximizes the use of available data by repeatedly dividing the dataset into training and validation sets.

The mean absolute percentage error (MAPE) is employed to assess the prediction accuracy for both training and testing data. MAPE is calculated as follows:

$$\text{MAPE} = \frac{1}{n}\sum_{i=1}^{n}\left|\frac{A_i - F_i}{A_i}\right| \times 100 \qquad (6)$$

where$A_i$ is the actual value and $F_i$ is the forecasted value. This metric provides a straightforward and interpretable measure of forecast accuracy [21].

**4. result**

### 4-1. feature importance

In machine learning prediction model, the predictors in dataset have essential role to improve the accuracy of model. The better the predictors can explain the behavior of the target variable, the more accurate the forecast will be. Instead of using research literature to select predictors, this study has used the large dataset of the World Bank to select the most effective predictor. Based on Goldani and Asadi [22], The Edit Distance on Real sequence (EDR) method was chosen for feature selection due to its lower sensitivity to sample size and simpler calculations. The importance of each predictor for each dataset is shown in fig1.

Fig2. Feature Importance

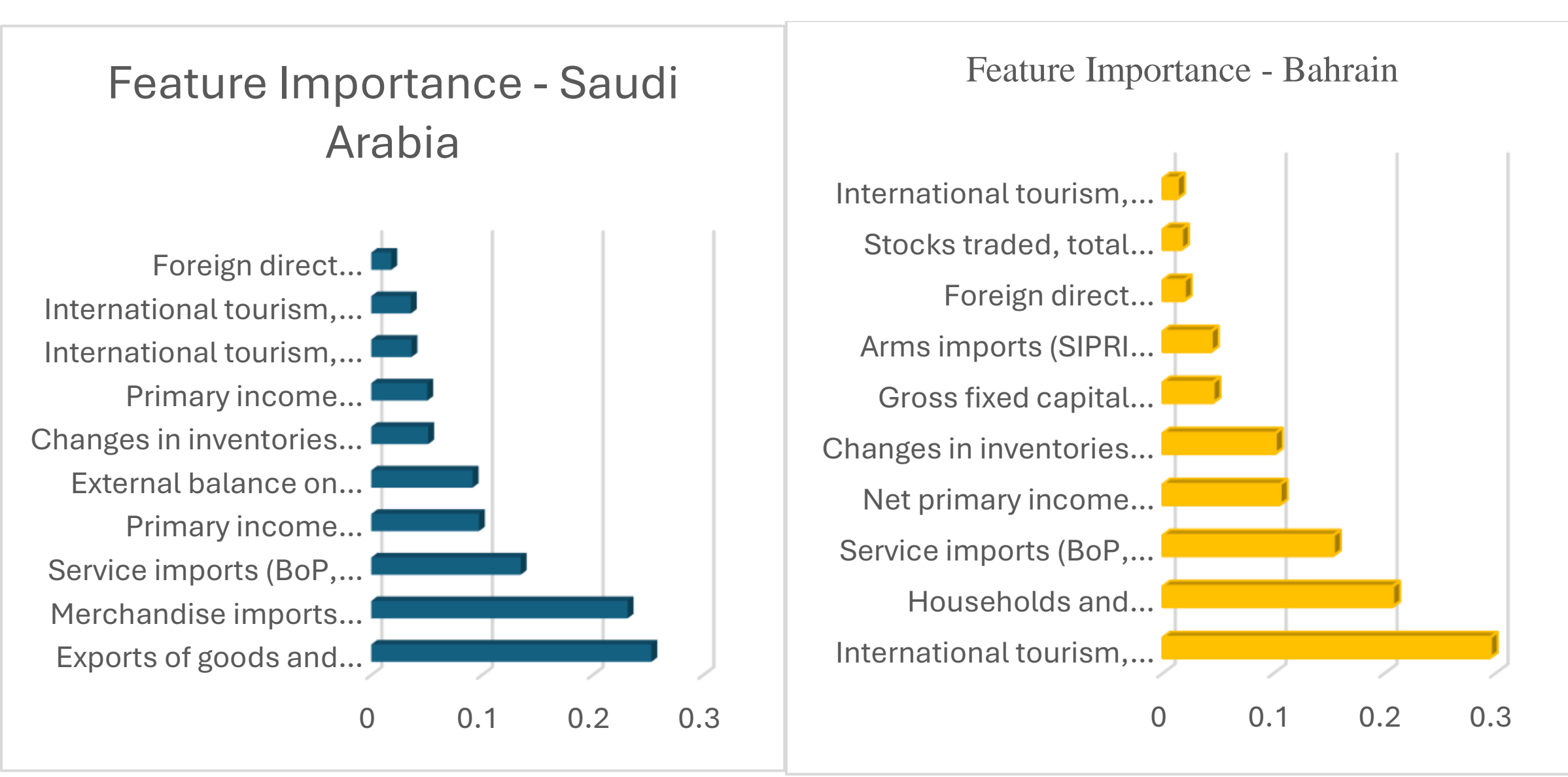

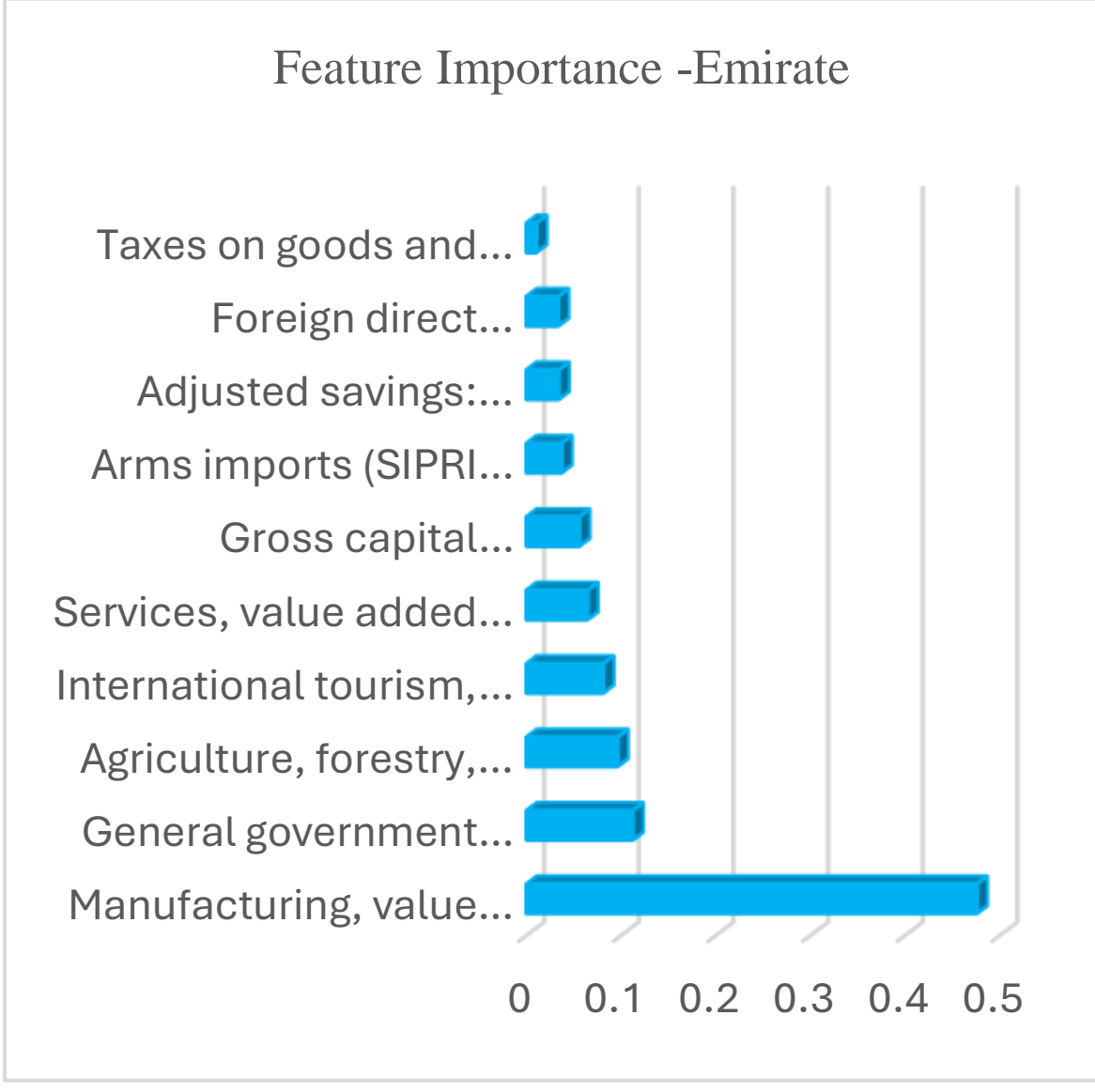

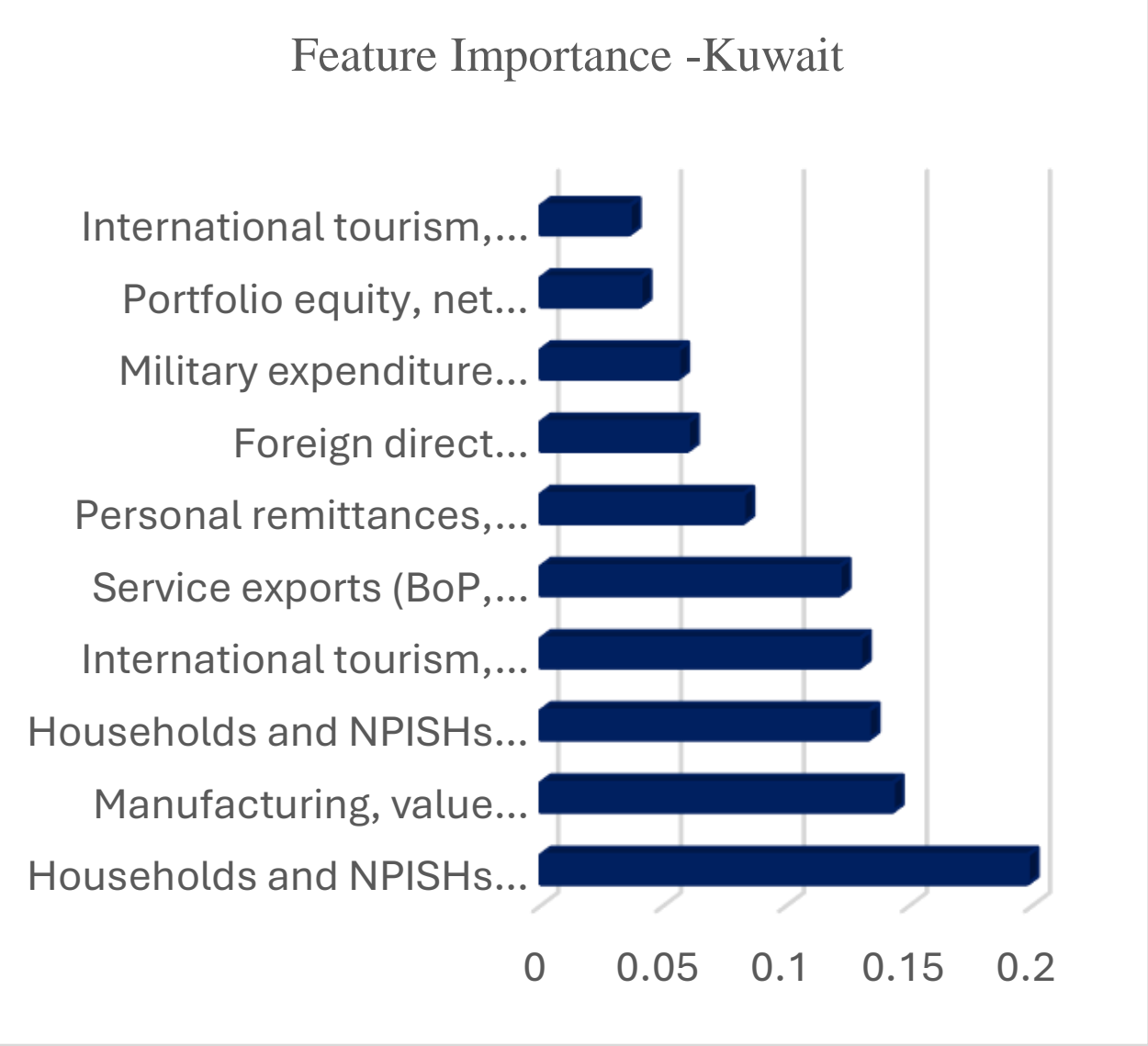

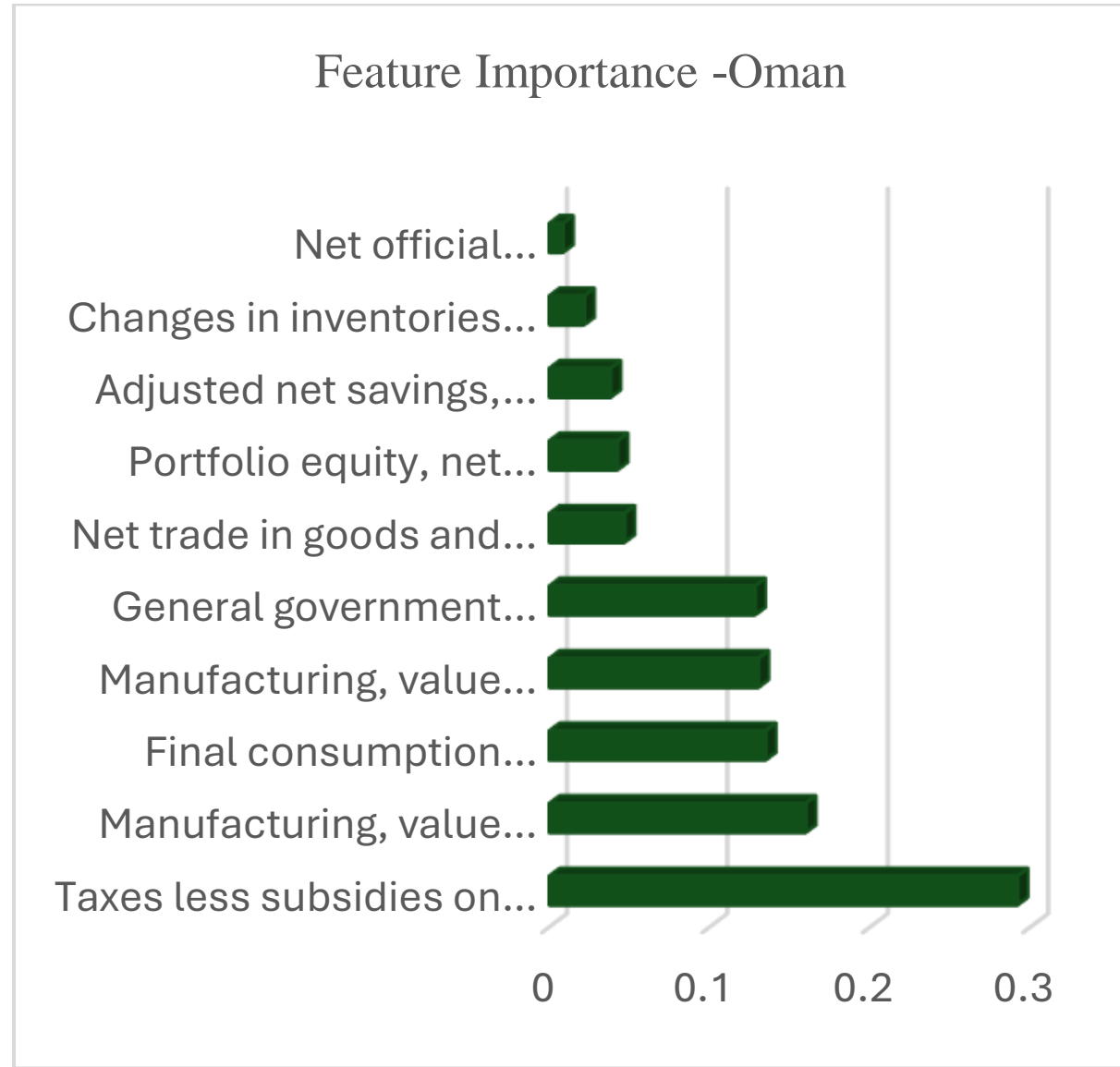

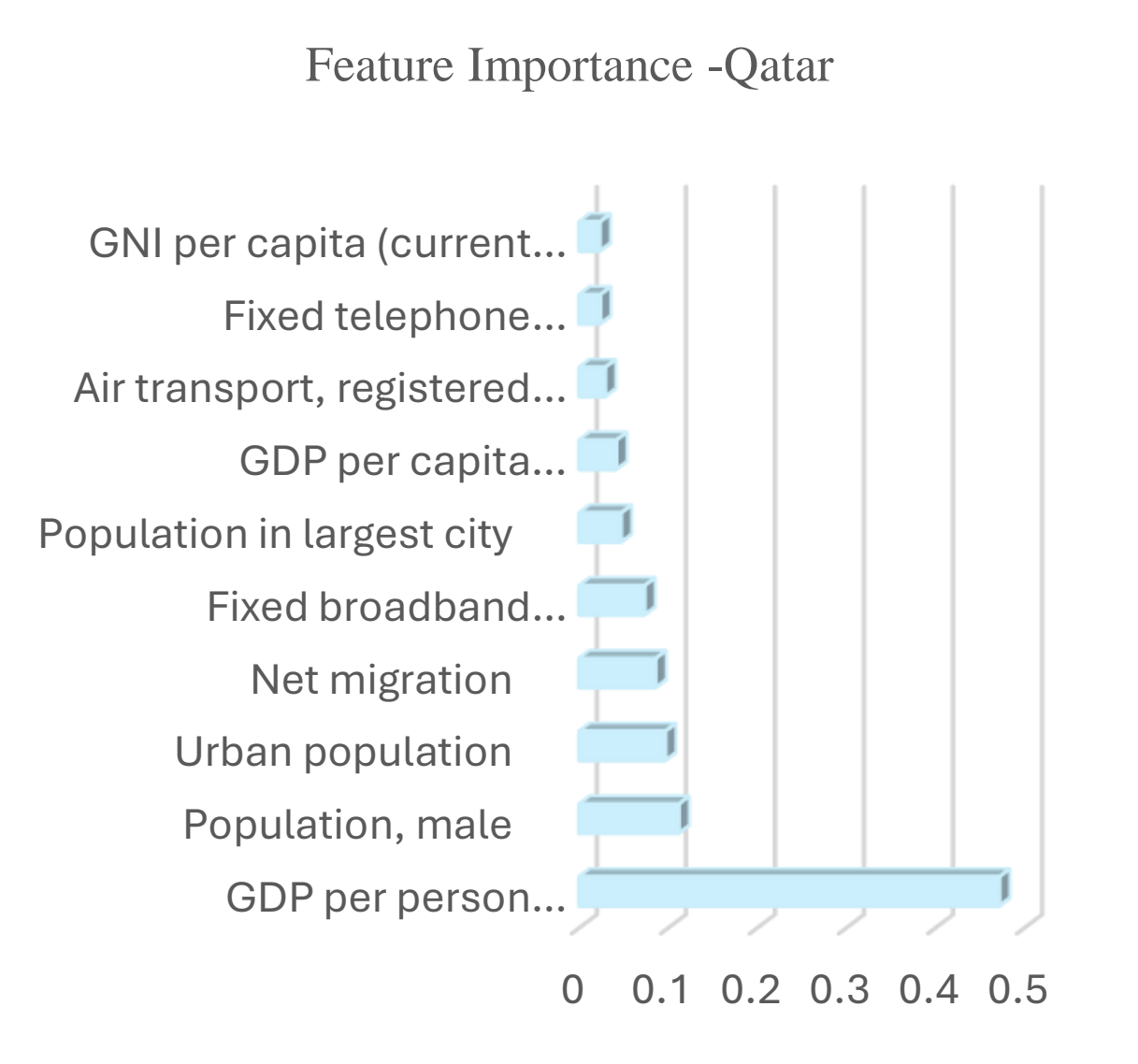

Based on the variables in the table, they can be broadly categorized into four main groups: Economic, Social, Cultural, and Military. For each country, economic variables seem to be primarily priority, followed by social indicators, with a smaller emphasis on cultural and military metrics. The majority of variables fall under the economic category, reflecting the emphasis on trade, investment, production, and consumption. Social variables come next in priority, focusing on indicators that impact the quality of life, infrastructure, and population metrics. The cultural factor was identified through tourism variables as one of the other dimensions affecting the political stability of a society.

**4-2. training and testing split**

The whole predictors and target value were split into training and tests to evaluate the performance of the model in each dataset. To improve the performance of the models, a model was built within the sample forecast. within the sample forecast used 1996 to 2018 to estimate the model. Using this model, the forecaster would then predict values for 2018-2023 and compare the forecasted values to the actual known values. Fig3 Compare the value of MAPE in both testing and training data set. Whole training MAPE are in Good forecasting. Between dataset, Saudi Arabia, Emirate and Qatar have good performance. Bahrain, Oman and kuwait have reasonable forecasting.

Fig3. MAPE of training and testing dataset

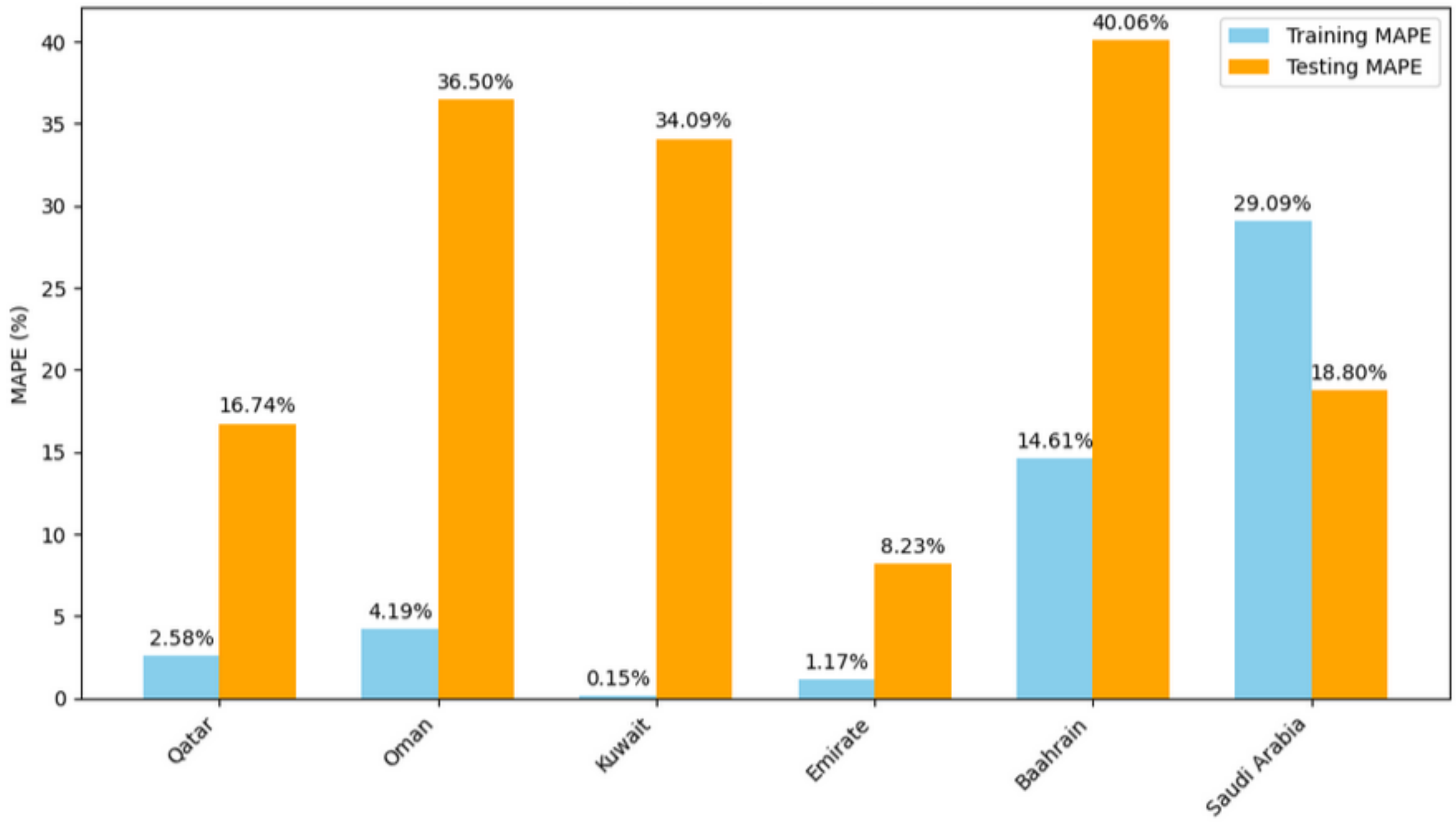

**4-3. simulation the predictors**

The next step is to predict the behavior of the political stability index in the six countries under investigation for the next five years. The predictors are simulated for the next five years. The Arima model was used to simulate 5 years of predictors. The historical trends of each time series were used to predict the future trend of them.

**4-4. forecast of political stability**

Fig. 4 shows the prediction model of each data set as well as the prediction of the political stability index for the six countries of Saudi Arabia, UAE, Kuwait, Oman, Bahrain and Qatar for the next five years. The first part of the figur predicts the training data along with the actual values, the second part predicts the testing values along with the actual values and finally the third part is the 5-year forecast of the political stability index for the mentioned countries based on the model.

Fig4. prediction model of each data set

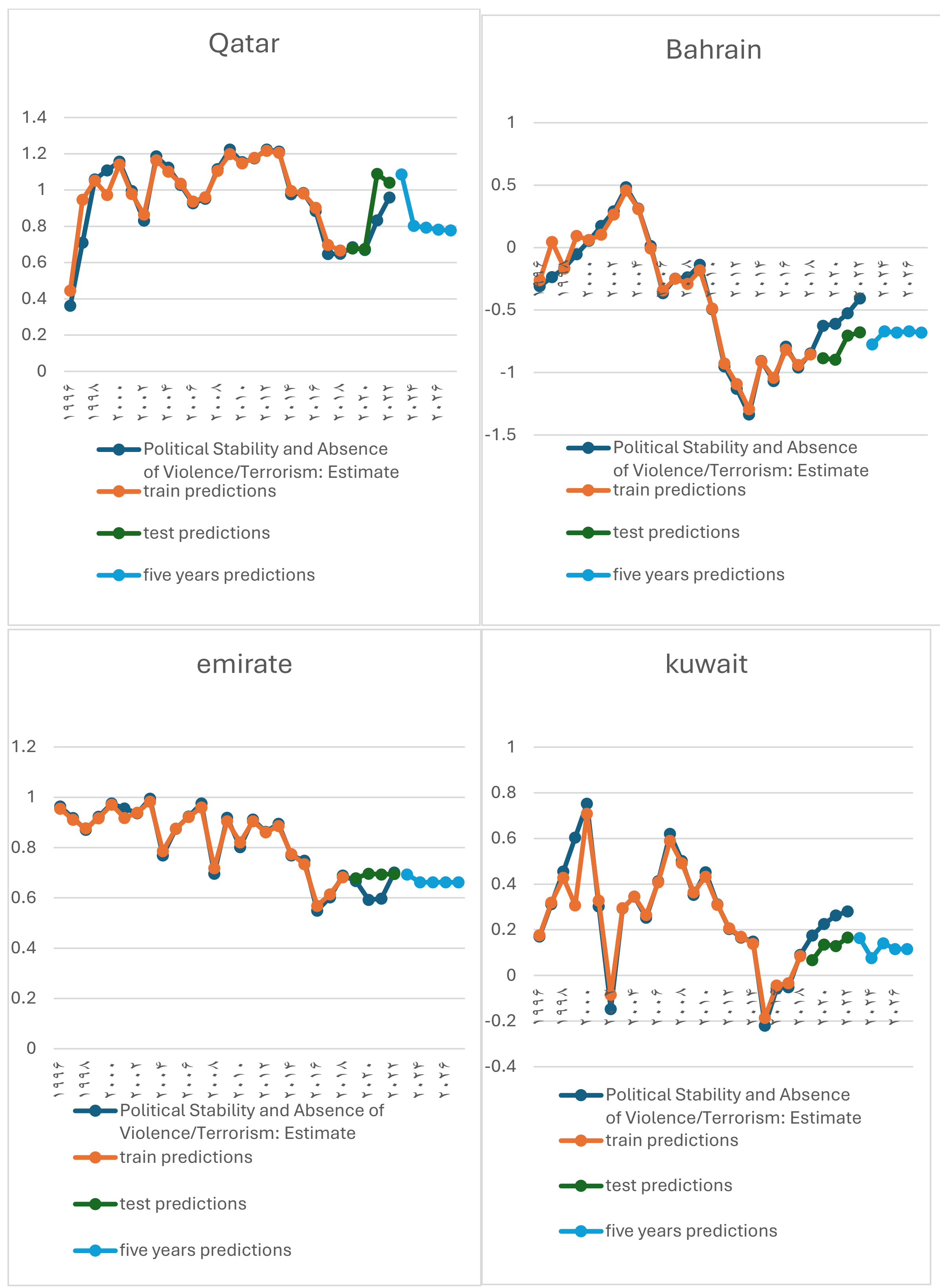

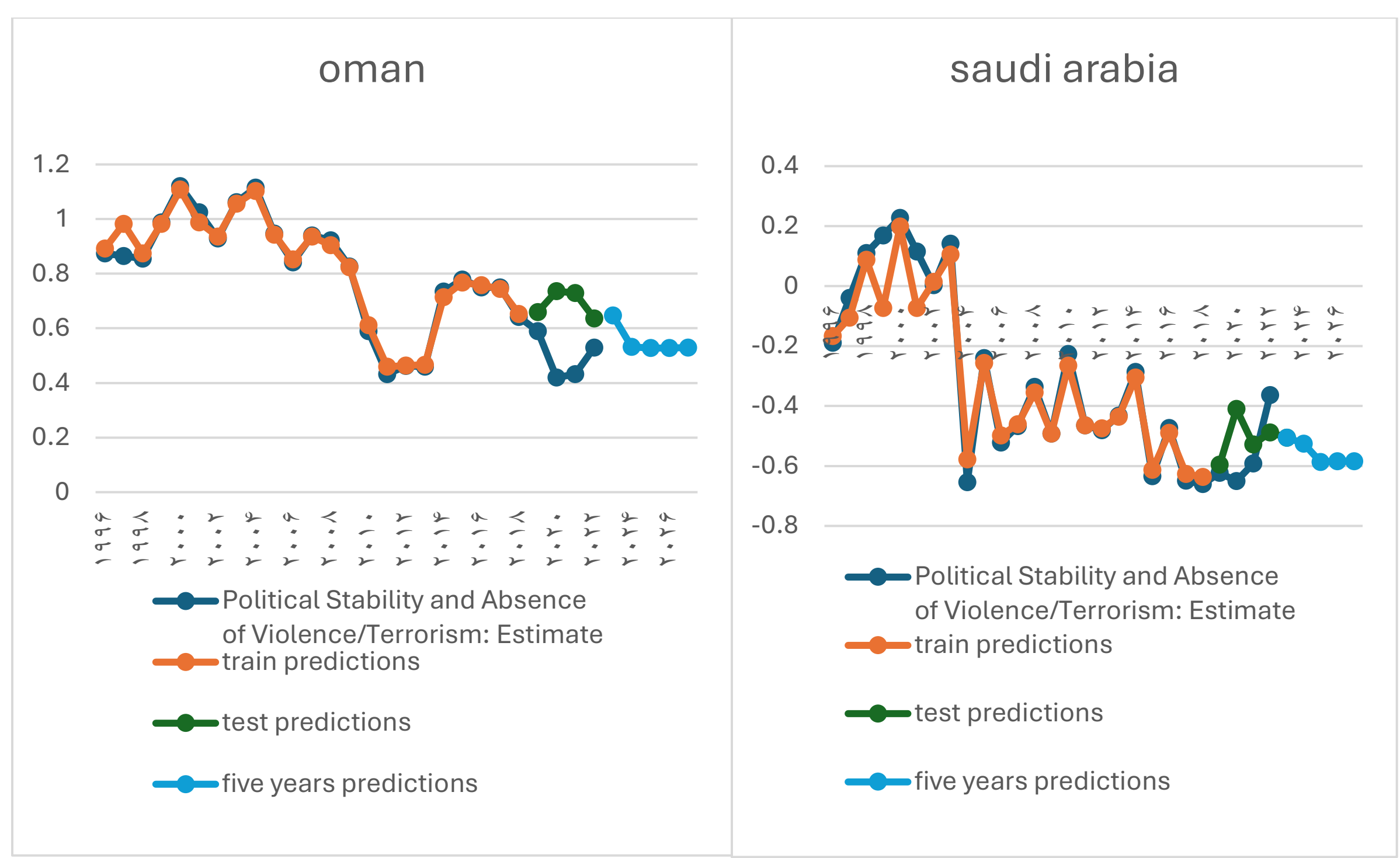

According to the forecasts, the three countries of Oman, the United Arab Emirates and Qatar will have better political stability conditions than the other three countries. Based on the historical data of political stability, the two countries of Saudi Arabia and Bahrain have a negative political stability index as before, which indicates the unfavorable conditions of political stability in these two countries. The ascending and descending status of the graphs shows that there will be no significant changes in political stability in almost all six countries, and the graphs are somewhat stable and will have slight changes during these five years.

Fig4. prediction of polotical stability index

1.086381 0.801757 0.792731 0.782011 0.777019
0.693578 0.662517 0.661938 0.662687 0.661921
0.532687 0.527431 0.527813 0.528734
0.163598 0.076147 0.140724 0.116193 0.116193
1 2 3 4 5
-0.505031 -0.52608 -0.586706 -0.584101 -0.584101
-0.77477 -0.670364 -0.679559 -0.669097 -0.679392
bahrain oman qatar emirate saudi arabia kuwait

## 5. Conclusion

This study has identified factors affecting political stability as well as predicting political stability for six GCC countries(Arabic Saudi Arabia, United Arab Emirates, Kuwait, Qatar, Oman, and Bahrain). By using the large dataset from the World Bank and employing the Edit Distance on Real sequence (EDR) method for feature selection, the most effective predictors were identified to improve the accuracy of the political stability forecasting model. The Arima model was applied to simulate future trends, with the model yielding reasonable forecasting accuracy based on MAPE values. The variables that had the most repetition in all six datasets are GDP, foreign investment, tourism variables, imports, military expenditures, etc.

These forecasts indicate relatively better political stability for Oman, the UAE, and Qatar, while Saudi Arabia and Bahrain will continue to retain negative political stability indices, depicting unfavorable political situations in these countries. The forecasted graphs for all the six countries indicate an overall stability with small fluctuations; this shows that there is no drastic change in the political stability of the various countries as forecast. These may also allow policymakers in those countries to be prepared for any challenge and implement strategies toward better governance and stability.

Based on the analysis and findings in this study, here are some targeted policy recommendations for the six GCC countries, aimed at enhancing and maintaining political stability:

- Economic Diversification Initiatives: Countries with persistent political instability indicators, like Saudi Arabia and Bahrain, should accelerate diversification efforts. Fostering industries like technology, renewable energy, manufacturing, and services (including tourism) can help balance economic dependencies.
- Strengthen Foreign Direct Investment (FDI) Policies: Oman, Qatar, and the UAE, showing relatively positive stability forecasts, can capitalize on this by enhancing FDI-friendly policies, which could serve as a stabilizing factor for the entire region.
- Focus on Social Stability through Employment Initiatives: Invest in education, skill development, and job creation, especially in Bahrain and Saudi Arabia, where indicators suggest a negative stability trend.
- Tourism Development Programs for Cultural and Economic Stability: With tourism showing as a strong variable, expanding this sector can offer both economic and social benefits. Oman, Qatar, and the UAE, with more favorable forecasts, can be role models for sustainable tourism practices.

By addressing the economic and social factors identified in this study, these policy recommendations aim to enhance political stability, facilitate sustainable growth, and prepare each country for potential future challenges.

**Refrence**